# Mössbauer spectroscopy evidence for the lack of iron magnetic moment in superconducting FeSe

A. Błachowski[1], K. Ruebenbauer[1*], J. Żukrowski[2], J. Przewoźnik[2], K. Wojciechowski[3], and Z. M. Stadnik[4]

[1]*Mössbauer Spectroscopy Division, Institute of Physics, Pedagogical University*
*PL-30-084 Cracow, ul. Podchorążych 2, Poland*

[2]*Solid State Physics Department, Faculty of Physics and Applied Computer Science,*
*AGH University of Science and Technology*
*PL-30-059 Cracow, Al. Mickiewicza 30, Poland*

[3]*Department of Inorganic Chemistry, Faculty of Material Science and Ceramics,*
*AGH University of Science and Technology*
*PL-30-059 Cracow, Al. Mickiewicza 30, Poland*

[4]*Department of Physics, University of Ottawa, Ottawa, Ontario K1N 6N5, Canada*

*Corresponding author: sfrueben@cyf-kr.edu.pl

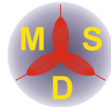

**Abstract**

Superconducting FeSe has been investigated by measurements of the magnetic susceptibility versus temperature and Mössbauer spectroscopy at various temperatures including strong external magnetic fields applied to the absorber. It was found that isomer shift exhibits sharply defined increase at about 105 K leading to the lowering of the electron density on iron nucleus by $0.02$ electron a.u.$^{-3}$. Above jump in the electron density is correlated with the transition from the *P4/nmm* to the *Cmma* structure, while decreasing temperature. Mössbauer measurements in the external magnetic field and for temperatures below transition to the superconducting state revealed null magnetic moment on iron atoms. Hence, the compound exhibits either Pauli paramagnetism or diamagnetic behavior. The principal component of the electric field gradient on the iron nucleus was found as negative on the iron site.

## 1. Introduction

Recent discovery of the iron-based superconductivity [1] resulted in the great scientific activity in similarity to the previous discovery of the high temperature superconductivity in cuprates [2]. Iron-based superconductors belong to pnictides or chalcogenides and these compounds have been already investigated in the past [3-5]. The simplest superconductor of this class belongs to the iron-selenium system, as it is simple binary system [6]. There are several compounds in the iron-selenium system [7], but superconductivity has been discovered for a compound being formed close to the FeSe stoichiometry. The following crystal phases could form close to the above composition range: 1. tetragonal *P4/nmm* structure called further β-phase, 2. hexagonal *P6$_3$/mmc* structure similar to the NiAs with various defects - called δ-phase and 3. hexagonal phase ($Fe_7Se_8$ formal composition) with two distinct kind of defect order, i.e., *3c* or *4c* [8]. The kind of order in this phase depends on the cooling rate from the high temperature [8]. All of the above phases order magnetically except tetragonal phase. The magnetic transition temperatures lie above room temperature except for the NiAs type δ-phase. The latter phase orders above liquid nitrogen temperature and the magnetic ordering process is accompanied by the lowering of the crystal symmetry [9]. A tetragonal phase transforms into *Cmma* orthorhombic phase in the broad temperature range centered about 90 K [10, 11] depending on the crystal size. Superconductivity has been discovered for this metallic phase and the transition temperature lies within the stability range of the orthorhombic phase *Cmma*. It seems to depend on the stoichiometry [12], i.e., it occurs for nearly stoichiometric compositions. It depends strongly on the applied pressure and dopants [13, 14]. An orthorhombic (and probably tetragonal) phase transforms to the hexagonal phase [14] for very high pressure. The latter phase seems to be metallic and non-magnetic like corresponding orthorhombic or tetragonal phase. Efforts to prepare samples containing pure tetragonal phase failed. It is always contaminated by some amount of hexagonal phases and by some amount of the metallic α-Fe. Usually traces of $Fe_3O_4$ occur as well.

There are some questions to be addressed for the tetragonal/orthorhombic FeSe system. The most important question is concerned with the electron spin density on iron. The Mössbauer spectroscopy seems particularly useful here, as iron makes the basic constituent of the compound. Mössbauer spectra obtained in the strong external magnetic field preferably below transition to the superconducting state are able to solve this problem. Another point is concerned with the eventual change of the electron (charge) density on iron nucleus during transition from *P4/nmm* to *Cmma* structure. It seems that superconductivity could be obtained solely in the *Cmma* phase, i.e., transition from *P4/nmm* into lower symmetry *Cmma* structure is essential to obtain superconductivity.

We report here on the preparation of the tetragonal phase, X-ray powder diffraction, susceptibility measurements and investigation by using Mössbauer spectroscopy.

## 2. Experimental

The sample was prepared by applying direct reaction between iron and selenium. Iron powder (AlfaAesar 99.5 % purity, 99.9+ % metal base) having spherical particles below 10 μm diameter was reduced under flow of 15 vol.% of $H_2$ and 85 vol.% of Ar mixture for one hour at 500 °C in the silica tube. A formal stoichiometry was chosen as $Fe_{1.05}$Se. Hence, proper amount of selenium (AlfaAesar 99.999+ % purity) in the form of shots having 2-4 mm size and reduced iron was sealed in the evacuated silica tube in order to prepare sample of 2 g. The tube was enclosed in another evacuated silica tube. A synthesis was carried at 750 °C for six days. Subsequently the sample was slowly cooled with the furnace to the room temperature. Resulting ingot was powdered in the mortar, powder sealed in the evacuated silica tube, and the sample was annealed at 420 °C for 48 hours and subsequently quenched in the ice water.

Powder X-ray diffraction pattern was obtained at room temperature by using Siemens D5000 diffractometer. The Cu-Kα radiation was used with the pyrolytic graphite monochromator on the detector side. Data were analyzed by means of the Rietveld method as implemented in the FULLPROF program [15].

Magnetic susceptibility was measured by means of the vibrating sample magnetometer (VSM) of the Quantum Design PPMS-9 system. The pelletized sample having mass 66.046 mg was cooled in the zero external magnetic field (no more than 0.7 Oe) to 2 K. It was subsequently warmed up to 355 K in the applied field of 5 Oe. Susceptibility was measured in the sweep mode in the field of 5 Oe during cooling sample having above history down to 2 K. Afterwards susceptibility was measured versus increasing temperature to 305 K in the field of 5 Oe.

Mössbauer spectra were collected in the temperature 4.2 K, in the range 75–120 K with the step 5 K and in the external magnetic field oriented along the γ-ray beam axis. Spectrum for the field 5 T was recorded with the source and absorber at 5.0(8) K, while the spectrum for the field 9 T was recorded with the source and absorber at 4.7(1) K. For the remaining measurements $^{57}$Co(Rh) source was kept at room temperature. Additionally spectra on the extended velocity range were collected at room temperature (RT) and at 80 K. All shifts are reported versus room temperature metallic iron. Data were analyzed by means of the MOSGRAF program [16].

**3. Results**

Figure 1 shows diffraction pattern. A tetragonal β-FeSe phase *P4/nmm* makes 93.7(6) % contribution. It has the following lattice constants $a$=3.7720(1) Å and $c$=5.5248(1) Å. The hexagonal phase makes 4.3(2) % contribution. It has been fitted as *P6$_3$/mmc* (δ-FeSe) disordered structure with the following lattice constants $a$=3.643(1) Å and $c$=5.909(2) Å. However Mössbauer data indicate that in fact it is one of the $Fe_7Se_8$ structures. Some traces of the magnetite ($a$=8.397(1) Å) and metallic iron ($a$=2.867(1) Å) were found as well.

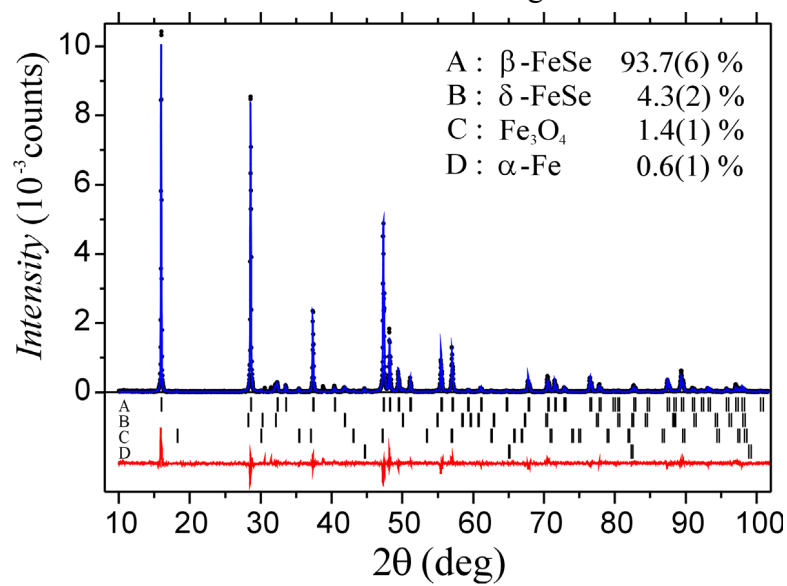

**Figure 1** (Color online) Powder X-ray diffraction pattern obtained at room temperature and by using Cu-Kα radiation. The pattern was obtained for the scattering angle 2θ varying between 10-158 deg.

A : β-FeSe 93.7(6) %
B : δ-FeSe 4.3(2) %
C : $Fe_3O_4$ 1.4(1) %
D : α-Fe 0.6(1) %

Mössbauer spectra obtained at room temperature (RT) and 80 K for high velocity ranges are shown in Figure 2 together with magnetic field distributions obtained for the magnetically ordered components by means of the Hesse-Rübartsch method [17]. The spectrum is dominated by the non-magnetic component originating in the tetragonal (RT) or orthorhombic phase (80 K) undergoing to the superconducting state at low temperatures. Field distribution clearly shows contribution from magnetite and one can even see Vervey transition upon cooling sample to 80 K. A contribution originating in the metallic iron is clearly seen as well. Additionally there are many small fields evolving to higher fields upon cooling with marked change in the distribution shape. They correspond to the fields expected for the $Fe_7Se_8$ structure including spin rotation upon cooling [18]. Hence, we conclude that the hexagonal phase is of the $Fe_7Se_8$ type (probably *3c* according to the thermal history of the sample). Therefore important minor phases are already magnetically ordered at room temperature with hyperfine fields not smaller than 10 T.

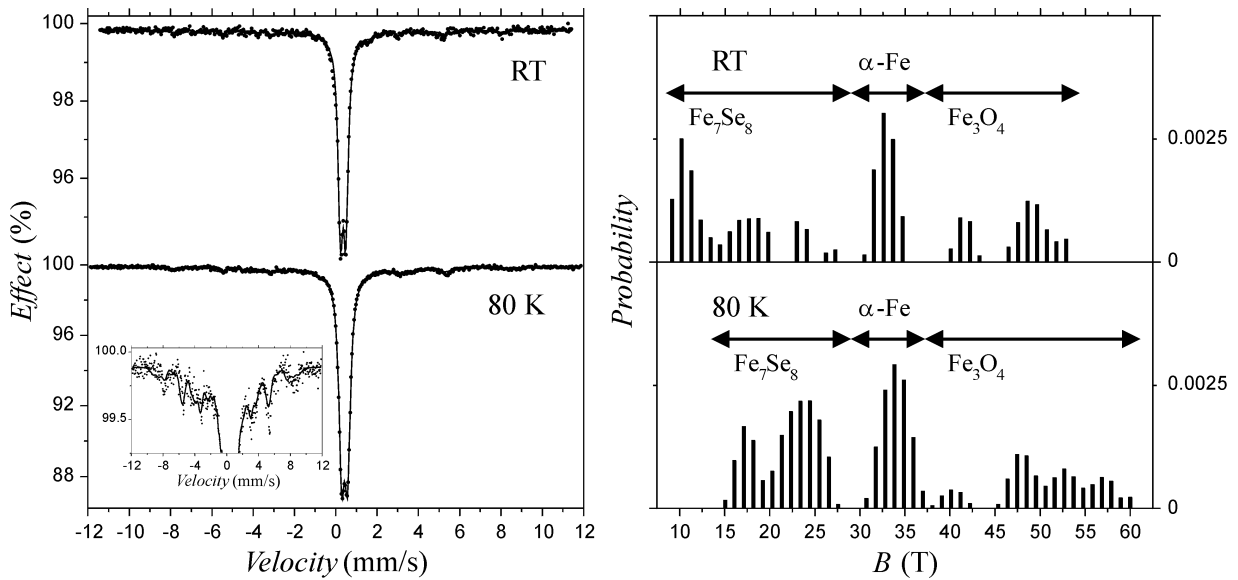

**Figure 2** Mössbauer spectra obtained for high velocity ranges together with the iron hyperfine field distributions in the minor phases. Non-magnetic contribution due to the dominant phase (tetragonal at RT and orthorhombic at 80 K) is fitted as doublet. Inset shows 80 K spectrum with the expanded vertical scale.

Results of the magnetic susceptibility measurements are shown in Figure 3. One can see spin rotation in the hexagonal phase at about 125 K [18]. Vervey transition in magnetite does not show up. Typical magnetic anomaly is seen at 80-105 K, and it correlates with the reported transition between orthorhombic and tetragonal phases [10, 11, 19]. Above transition occurs at about 105 K in the case of polycrystalline material as ours [19]. Above anomaly is much weaker (or absent) in single crystals and the transition temperature is lowered [19]. Discernible anomaly has been observed in the polycrystalline material by various groups [6, 20, 21].

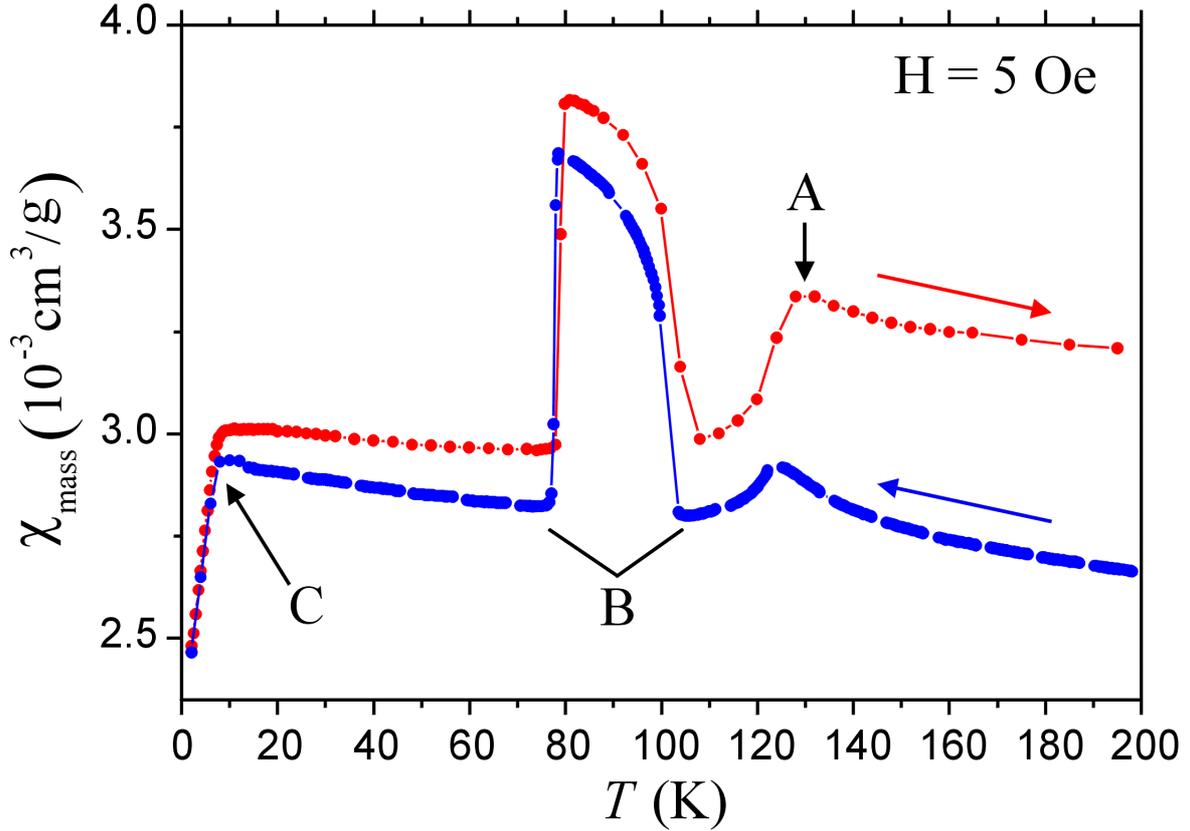

**Figure 3** (Color online) Susceptibility measured upon cooling and subsequent warming of the sample in the field of 5 Oe is shown. The point A corresponds to the spin rotation in the hexagonal phase, region B shows typical magnetic anomaly for this system, while point C shows transition to the superconducting state.

Figure 4 shows selected Mössbauer spectra measured for low velocity ranges at various temperatures. Spectra were fitted within transmission integral approximation by quadrupole split doublet. Essential fit results are summarized in Table 1. One has to realize that in the *P4/nmm* and *Cmma* structures there is single crystallographic regular iron site per unit cell [11], i.e., all iron sites are equivalent within the chemical unit cell.

Total spectral shift versus temperature is shown in Figure 5. One can see distinct jump of the isomer shift by $+0.006(1)$ mm/s at about 105 K superimposed on the continuous change due the second order Doppler shift (SOD). This jump corresponds to the lowering of the electron density on the iron nucleus in the *Cmma* phase by 0.02 electron a.u.$^{-3}$ [22] in comparison with the *P4/nmm* phase at the same temperature. The jump of the isomer shift correlates with the onset of the magnetic anomaly and transition between tetragonal and orthorhombic phase. Other hyperfine parameters remain constant in this temperature range. In particular the quadrupole coupling constant does not change. It means that the local arrangement of Se atoms around Fe atom does not change during the phase transition. Spectral lines are slightly broadened indicating some kind of disorder in both phases.

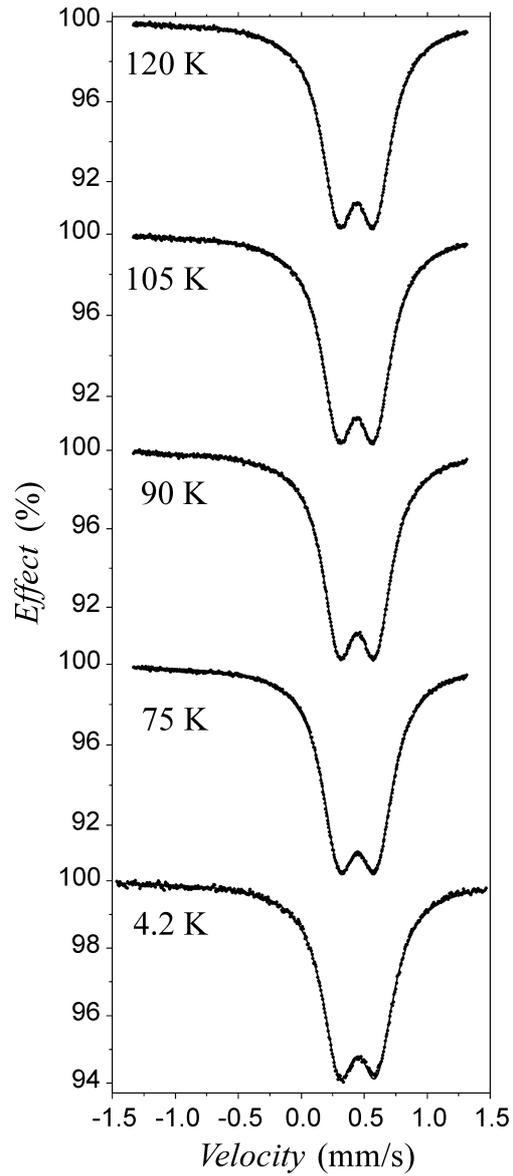

**Figure 4** Mössbauer spectra measured for low velocity ranges.

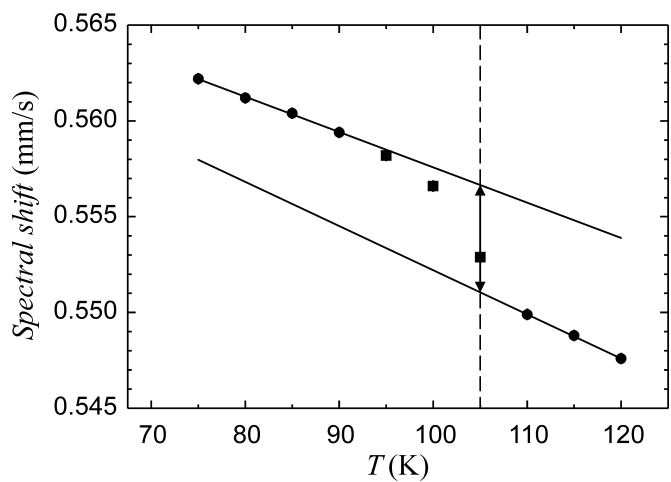

**Figure 5** Spectral shift versus temperature in the phase transition region. Points marked by rectangles are excluded from respective linear regression fits.

**Table 1**
Essential results obtained by using Mössbauer spectroscopy. The spectra were processed within the transmission integral approximation. The symbol *S* denotes total spectral shift versus RT metallic iron. The symbol Δ stands for the quadrupole splitting, while the symbol Γ denotes absorber common line width.

| *T* (K) | *S* (mm/s) | Δ (mm/s) | Γ (mm/s) |
|---|---|---|---|
| 120 | 0.5476(3) | 0.287(1) | 0.206(1) |
| 105 | 0.5529(3) | 0.287(1) | 0.203(1) |
| 90 | 0.5594(3) | 0.286(1) | 0.198(1) |
| 75 | 0.5622(3) | 0.287(1) | 0.211(1) |
| 4.2 | 0.5640(4) | 0.295(1) | 0.222(1) |

Figure 6 shows spectra obtained in the external magnetic fields of 5.0 T and 9.0 T and below superconducting transition temperature. Spectra were fitted within transmission integral approximation averaging over direction between the magnetic field and principal axis of the electric field gradient (EFG) due to the powder form of the absorber [16] - in one-degree steps. External field is much higher than the first critical field and practically all iron nuclei experience applied field. Data evaluation reveals that hyperfine fields are exactly equal to the applied field, i.e., 5.00(1) T and 8.96(1) T, respectively. Hence, there is no magnetic moment on the iron nuclei.

We have assumed that EFG is axially symmetric according to the local symmetry. We have found negative value for the principal component of the EFG.

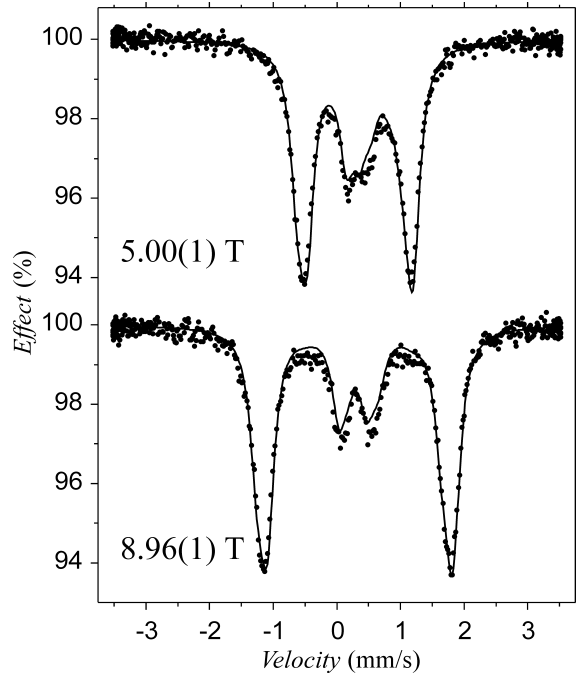

**Figure 6** Mössbauer spectra obtained in the external magnetic field aligned with the beam. Spectrum for the field 5.0 T was recorded with the source and absorber at 5.0(8) K, while the spectrum for the field 9.0 T was recorded with the source and absorber at 4.7(1) K. Field values shown in Figure are obtained from the fit of spectra.

## 4. Conclusions

There are two basic conclusions obtained upon having performed above investigations:

1. There is no measurable magnetic moment on the iron atoms in the superconducting FeSe.
2. The electron density on iron nucleus lowers by 0.02 electron a.u.$^{-3}$ during transition from tetragonal to the orthorhombic phase.

Similar results (no magnetic moment on iron) were obtained by the Mössbauer measurements in the external magnetic field for LaFeAsO$_{0.89}$F$_{0.11}$ [23], La$_{0.87}$Ca$_{0.13}$FePO [24], LaFeAsO$_{0.93}$F$_{0.07}$ [24] and other iron-based superconductors like filled skutterudites (LaFe$_4$P$_{12}$) [25]. Mössbauer measurements on the metallic-like Fe$_{1.04}$Se sample in the 5 T field at 4.2 K did not show any appreciable contribution due to the internal hyperfine magnetic field on iron [3]. Similar results were obtained for non-stoichiometric insulator Fe$_{1+x}$S [26] and many other low-spin ferrous and insulating compounds like cyanides. The situation is somewhat different for the superconducting high-pressure hexagonal ε-Fe [27], where the external field being able to destroy superconductivity induces a hyperfine field on the iron nucleus far beyond the limit of the Knight shift [28]. The Knight shift for the iron-based superconductors investigated up to now by $^{57}$Fe (14.41-keV line) Mössbauer spectroscopy is below detection limit by this method.

Therefore it seems that iron-based superconductors are similar to cuprates. One needs non-magnetic electron configuration on iron and the metallic environment to have a chance to observe superconductivity. Iron is very similar to copper in such situation.